\documentclass[lettersize,journal, 10pt]{IEEEtran}
\usepackage{amsmath,amsfonts}
\usepackage{algorithmic}
\usepackage{algorithm}
\usepackage{amssymb}
\usepackage{array}
\usepackage{multirow}
\usepackage[belowskip=-0pt,aboveskip=0pt,font=footnotesize]{caption}
\usepackage{textcomp}
\usepackage{stfloats}
\usepackage{url}
\usepackage{verbatim}
\usepackage{graphicx}
\usepackage{cite}
\usepackage[dvipsnames]{xcolor}
\usepackage{soul}
\usepackage{subfigure}
\usepackage{soul}
\usepackage[normalem]{ulem}

\begin{document}
\pagenumbering{arabic}
\title{GAN-RXA: A Practical Scalable Solution to Receiver-Agnostic Transmitter Fingerprinting}

\author{
\IEEEauthorblockN{
Tianyi Zhao,
Shamik Sarkar,~\IEEEmembership{Member,~IEEE,}
Enes Krijestorac, 
Danijela Cabric,~\IEEEmembership{Fellow,~IEEE}}

\thanks{This work was supported by SpectrumX, which is an NSF Spectrum Innovation Center funded via Award 2132700.}
\thanks{Tianyi Zhao, Enes Krijestorac and Danijela Cabric are with the Electrical and Computer Engineering Department,
University of California at Los Angeles, Los Angeles, CA 90095 USA (e-mail:
zhaotianyi@ucla.edu; enesk@ucla.edu;  danijela@ee.ucla.edu).}
\thanks{Shamik Sarkar was with the Electrical and Computer Engineering Department, University of California, Los Angeles, CA 90095, USA. He is now with the Department of Electronics and Communications Engineering at Indraprastha Institute of Information Technology Delhi (IIIT-Delhi), Delhi 110020, India (e-mail: shamik@iiitd.ac.in).}
}

\maketitle

\begin{abstract}
Radio frequency fingerprinting has been proposed for device identification. However, experimental studies also demonstrated its sensitivity to deployment changes. Recent works have addressed channel impacts by developing robust algorithms accounting for time and location variability, but the impacts of receiver impairments on transmitter fingerprints are yet to be solved. In this work, we investigate the receiver-agnostic transmitter fingerprinting problem, and propose a novel two-stage supervised learning framework (RXA) to address it. In the first stage, our approach calibrates a receiver-agnostic transmitter feature-extractor. We also propose two deep-learning approaches (SD-RXA and GAN-RXA) in this first stage to improve the receiver-agnostic property of the RXA framework. In the second stage, the calibrated feature-extractor is utilized to train a transmitter classifier with only one receiver. We evaluate the proposed approaches on the transmitter identification problem using a large-scale WiFi dataset. We show that when a trained transmitter-classifier is deployed on new receivers, the RXA framework can improve the classification accuracy by $19.5\%$, and the outlier detection ROC AUC by $12.0\%$ compared to a naive approach without calibration. Moreover, GAN-RXA can further increase the closed-set classification accuracy by $5.0\%$, and the outlier detection ROC AUC by $12.5\%$ compared to the RXA approach.
\end{abstract}


\section{Introduction}
\subsection{Motivation}\label{motivation}
In wireless communication, a transmitter encodes its messages into electrical signals, converts them to electromagnetic signals (commonly radio frequency signals), and transmits over the air. The receiver captures the radio frequency (RF) signals, performs the same operations as the transmitter in reverse order, and extracts the original message. In this process, the RF signals are distorted due to the wireless channel and transceiver hardware imperfections. In other words, transmitters, channels, and receivers can implant their fingerprints in RF signals. 
Hence, these fingerprints are commonly known as `RF fingerprints.' While these RF fingerprints are unintentionally embedded in the signals, they can also reveal additional information about the transceivers or the radio environment. For example, RF fingerprints due to transmitter power amplifier non-linearity can reveal the identity of the transmitter \cite{p33}. Here, the transmitter's identity is the relevant additional information. Similarly, RF fingerprints due to the wireless channel can reveal the location of a transmitter \cite{p53}. Here, the transmitter's location is the relevant additional information about the radio environment. The goal of \emph{RF fingerprinting} is to extract the RF fingerprints and leverage the additionally embedded information for different applications.

Prior works have demonstrated the viability of leveraging RF fingerprints for a variety of applications \cite{p20}. For example, channel induced fingerprints can be utilized for applications such as localization \cite{p53}, channel prediction \cite{p14}, and radio map reconstruction \cite{p27}. In another direction, researchers have exploited transmitter fingerprints to perform device identification \cite{p54, p15, p17, p29, p33, p36, p47, p51, p1}, which is also our focus in this paper. Transmitter fingerprints typically come from radio hardware impairments as a result of manufacturing nuances and are inevitable due to non-ideal physical components. Consequently, each transmitter has its own unique fingerprints. The idea of wireless device identification using transmitter fingerprints, which is a physical layer authentication (PLA) technique, has the advantage of being robust against spoofing attacks \cite{p29} while saving computational overheads \cite{p30}.

Device identification using transmitter fingerprints, which we call \emph{transmitter fingerprinting}, can be performed in two different settings: closed-set and open-set. In a closed-set setting, the problem is to identify a transmitter among a known set of transmitters based on the captured RF signals.
On the other hand, in an open-set setting, the additional problem is to identify whether the captured RF signal originates from a known or unknown transmitter. In both settings, the transmitter fingerprinting problem can be framed as a classification task.
The classification must be done based on features that accurately capture the unique fingerprints of the transmitters.
These features are either explicitly or implicitly extracted from the received RF signals.
Thus, reliable classification requires robust transmitter feature extraction.

As RF signals also suffer from channel and receiver fingerprints, it is reasonable to consider their impacts on the extracted transmitter features. In fact, transmitter fingerprinting is sensitive to deployment variability as a result of variable channels and different receivers. These variations can lead to significant degradation in the transmitter-classification performance \cite{p10, p11, p26, p22, p31}. Recent studies have proposed channel-agnostic transmitter fingerprinting \cite{p5, p12}, and these techniques have demonstrated the ability to suppress channel fingerprints. However, in the context of transmitter fingerprinting, the problem of suppressing receiver fingerprints has not been sufficiently investigated. This is a crucial problem as receivers are often replaced due to malfunctioning or upgrades, and different receivers will have distinct receiver fingerprints. Thus, to make transmitter fingerprinting practical, it is necessary to provide receiver-agnostic solutions for transmitter fingerprinting. 

Some existing works address the receiver-agnostic transmitter fingerprinting problem partially in \cite{p47}, \cite{p19}, and \cite{p46}. For example, the approach in \cite{p46} requires extensive signal collection in the training stage for every new deployment. Similarly, the approach in \cite{p19} requires additional training in the testing stage. The approach in \cite{p47} requires the collaboration of multiple receivers in the testing stage. In contrast, in this work, we address the receiver-agnostic transmitter fingerprinting problem with a more practical and scalable deep learning approach. Specifically, our approach only
requires minimal signal collection on only one receiver for training in each new deployment (we call this attribute of our solution \emph{practical}), and accurate transmitter fingerprinting can be done on a new receiver without any additional retraining or tuning (we call this attribute of our solution \emph{scalable}).

\subsection{Related Work}
As discussed in the previous section, transmitter fingerprinting involves two steps, transmitter feature extraction and classification. Traditional approaches focus on the design of handcrafted features for transmitters and then perform the classification using statistical methods such as the likelihood ratio test. Additionally, in recent years, deep learning has been widely applied in transmitter fingerprinting and its efficacy has been demonstrated in both feature extraction and classification tasks.

\subsubsection{Handcrafted Feature Extraction Approaches}
These approaches rely on models of transmitter impairments and analyze the physical components in the circuits contributing to the transmitter fingerprints for extracting features. 
For example, prior works have shown that transmitters can be identified by fingerprints due to power amplifiers (PAs) and mixers. The authors in \cite{p33} model the nonlinear characteristics in PAs with Volterra series representation and use it as the transmitter features. The authors in \cite{p36} exploit the mixer imperfection and model the drifting oscillator offsets to use it as transmitter features. While individual physical components can provide their distinct fingerprints, it is natural to consider their collective contributions, which can form more comprehensive fingerprints for the different transmitters. Hence, more advanced feature representations for transmitter identification have been proposed to utilize the collective fingerprints better. For example, the authors in \cite{p17} present a Differential Constellation Trace Figure (DCTF) as the extracted features of a transmitter, which can retain IQ imbalance and carrier frequency offset (CFO). The authors in \cite{p51} present CepH, a feature representation exploiting the unintentional modulation of the pulse (UMOP) based on cepstrum analysis of the RF signals. As the feature representations become more complicated, the extraction of handcrafted features depends on a comprehensive analysis of the received signals. To deal with this challenge, deep learning approaches can be used as effective tools since they can save the efforts in handcrafted feature extraction and learn directly from the raw data without any complicated preprocessing.

\subsubsection{Deep Learning Approaches}
As deep learning typically has the capability of learning complicated models, it is hypothesized that deep learning can extract transmitter features without an explicit impairment model. Several works have demonstrated the validity of this hypothesis. At the same time, deep learning can also be applied in the classification stage when the extracted features are complicated. For example, Convolutional Neural Network (CNN) is applied in \cite{p15} and \cite{p54}, and the proposed model can directly extract features from IQ samples and classify the transmitters with minimal pre-processing. At the same time, CNN is also applied in \cite{p17} to classify the generated DCTFs and thus their corresponding transmitters.

\subsubsection{Channel-Agnostic Transmitter Fingerprinting}
In realistic settings, the task of transmitter fingerprinting must overcome the challenges of channel variations and receiver fingerprints. Studies have been conducted on channel-agnostic transmitter fingerprinting. The authors in \cite{p5} present ChaRRNets, a complex-valued CNN-based architecture for RF fingerprinting, which is robust to channel variations. ChaRRNets first applies Short Time Fourier Transform (STFT) on the signals and obtain spectra. Then, it sends the spectra into a channel-equivariant convolution layer followed by an invariant distance layer and obtains a channel-robust representation of the signals. Finally, it feeds the channel-robust representations into a 1-Dimensional (1D) real-valued CNN, which performs the classification. The authors in \cite{p12} also present a deep-learning based framework, together with a channel-independent spectrogram as the representation of received RF signals. The channel-independent spectrogram is generated by removing channel effect on STFT with the assumption that the channel remains constant within short time. Then, the spectrogram is fed into a CNN with triplet loss to obtain pairwise Euclidean distances. Finally, classification is completed by the K Nearest Neighbor (KNN) algorithm.

\subsubsection{Receiver-Agnostic Transmitter Fingerprinting}
As the channel fingerprints can be suppressed by the aforementioned techniques, the remaining problem is to address the receiver fingerprint, which has not been sufficiently investigated. Some works have studied the problem of receiver effects in the transmitter fingerprinting task. The authors in \cite{p19} propose to calibrate each receiver with a neural network-based transformation, to address the substantial classification performance degradation due to different receivers used for training and testing. This calibration method is efficient in classifier training, as it allows effective training with only a single receiver. However, each newly deployed receiver still needs to collect new signals and learn its own transformation function. A crowdsourced measurement method is presented in \cite{p47}, which utilizes a collective set of receivers to identify transmitters during testing. This method also partially solves the receiver fingerprint problem, because it may not be applicable under scenarios where multiple receivers are not available during the testing stage. The authors in \cite{p46} propose an algorithm based on adversarial training to address the receiver fingerprint problem. This algorithm enables effective transmitter classification on newly deployed receivers without retraining. However, it still requires extensive signal collection on multiple receivers in each new deployment during the training stage. In summary, existing works each solve the receiver fingerprint problem partially, with practical limitations on either training or testing stages. In this paper, we propose to solve the problem in a practical and scalable manner. The proposed approach does not require extensive signal collection for training in each new deployment or retraining on any new receiver. 

\begin{table*}
\caption{Comparison between our proposed GAN-RXA and some existing works on device authentication using deep learning.}
\centering
  \begin{tabular}{|c|c|c|c|}
  \hline
  Work & Category & Evaluations & Attributes\\
  \hline
  \multirow{2}{*}{\cite{p33, p36, p17, p51}} & \multirow{2}{*}{Handcrafted Features} & Simulation/ & Design handcrafted features \\
  & & QAM/ZigBee & extracted from RF signals \\
  \hline
  \multirow{2}{*}{\cite{p15, p54, p17}} & Deep Learning for & \multirow{2}{*}{WiFi/ZigBee} & Design deep learning approaches for automatic \\
  & Feature Extraction/Classification& & feature extraction and transmitter identification \\
  \hline
  \multirow{2}{*}{\cite{p5, p12}} & \multirow{2}{*}{Channel-Agnostic} & \multirow{2}{*}{WiFi/LoRa} & Address channel variations \\
  & & & between training and testing stages\\
  \hline
  \multirow{3}{*}{\cite{p19, p47, p46}} & \multirow{3}{*}{Receiver-Agnostic} & ZigBee/ & Address receiver variations  \\
  & & QAM/ & between training and testing stages\\
  & & LoRa & but with practicality or scalability limitations \footnotemark\\
  \hline
  \multirow{2}{*}{GAN-RXA} & \multirow{3}{*}{Receiver-Agnostic} &\multirow{3}{*}{WiFi} & Address receiver variations\\
  \multirow{2}{*}{(Our Work)} & & & between training and testing stages\\
  & & & in a practical and scalable manner\\
  \hline
  \end{tabular}
  \label{ratf_work}
\end{table*}
\footnotetext{Refer to Section \ref{motivation} for the notion of practicality and scalability.}
To summarize this section, in Table~\ref{ratf_work}, we contrast our work with relevant works concisely.

\subsection{Contributions}

Our contributions can be summarized as follows:

\begin{itemize}
\item We formulate the receiver-agnostic transmitter-fingerprinting (RATF) problem, as to build a practically scalable transmitter-authenticator that can be easily deployed in new environments and easily transferred to different receivers. We stress that besides classification accuracy, the practicality and cost of scalability of the solution are the major concerns in realistic scenarios.
\item We propose a novel two-stage supervised learning framework (RXA) to address the aforementioned RATF problem. In the first stage, our approach calibrates a transmitter feature extractor, which is robust against receiver fingerprints, in a controlled environment. Then, in the second stage, this feature extractor is applied to a new receiver that aims to authenticate a set of unseen transmitters in a targeted field. We use a deep learning based classifier for the authentication, which can be applied in both closed-set and open-set scenarios. We claim that this two-stage framework is able to address the RATF problem in a practical and scalable manner. With this approach, only one receiver is required to collect training data from different transmitters in the second stage, and each new receiver is able to classify the targeted transmitters without any further signal collection or retraining. 
\item We develop two deep learning based approaches for the receiver agnostic transmitter feature extractor (SD-RXA and GAN-RXA) in the first learning stage. We design the loss functions and deep neural network architectures so that the feature extractors are able to distinguish among transmitters while being insensitive to receiver differences. We also discuss the trade-offs between our two proposed approaches.
\item We evaluate our proposed approaches with extensive experiments on WiSig, a large-scale WiFi dataset \cite{p4} captured over the air. We show that our proposed framework can improve both the closed-set and open-set problem performances. Specifically, our proposed GAN-RXA approach can improve closed-set transmitter classification accuracy by $24.5\%$ and open-set transmitter outlier detection receiver operating characteristic (ROC) area under the curve (AUC) by $26.0\%$, compared to a naive approach without calibration, where a classifier trained on a single receiver is applied on another receiver.
\end{itemize}

The rest of the paper is organized as follows. Section \ref{pf} formulates the RATF problem. Section \ref{ap} explains the proposed approaches to solve the RATF problem. Section \ref{ds} discusses the dataset and the pre-processing of the data used in the experiments. Section \ref{ee} presents the experimental setup and results for the proposed approaches. Finally, section \ref{conclusion} concludes the paper. The acronyms used in this paper are summarized in Table \ref{acronyms}.

\begin{table}
\caption{Acronyms used in the paper}
\centering
  \begin{tabular}{|c|c|}
  \hline
  Acronym & Full Form \\
  \hline
  RATF & Receiver-agnostic transmitter fingerprinting\\
  \hline
  GAN& Generative Adversarial Network \\
  \hline
  RXA & Proposed two-stage RATF learning framework\\
  \hline
  SD-RXA & Proposed statistical distance based RXA framework\\
  \hline
  GAN-RXA & Proposed GAN based RXA framework\\
  \hline
  FE & Feature-extractor\\
  \hline
  ROC & Receiver operating characteristic\\
  \hline
  AUC & Area under the curve\\
  \hline
  \end{tabular}
  \label{acronyms}
\end{table}

\section{Problem Formulation} \label{pf}

In this section, we first discuss the RATF problem statement together with the essential concerns in solving it, and then we define the necessary metrics in evaluating a solution.

First, we consider the following model for received RF signals: assuming a perfectly generated signal $x$, the received signal $y$ is modeled as
\begin{equation}
y=f_{R_j}(f_{H_c}(f_{T_i}(x)))
\label{signal_model}
\end{equation}
where $f_{T_i}$ represents the unique fingerprint of a transmitter $T_i$, $f_{H_c}$ represents the channel fingerprint of a random channel $H_c$, and $f_{R_j}$ represents the unique fingerprint of a receiver $R_j$.

\begin{figure}[h]
    \centering
    \includegraphics[width=0.45\textwidth]{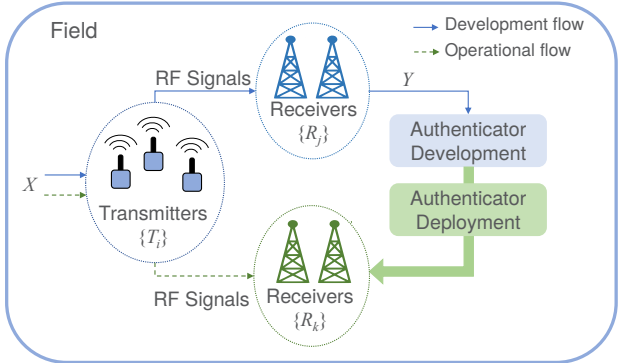}
    \caption{A demonstration of a field, where a solid blue arrow represents the flow of RF signals, a solid green arrow represents the transfer of a build authenticator, and a dashed green arrow represents possible RF signal transmssions. In each field, the building process of an authenticator needs to be practical, and the transfer of a built authenticator across different receivers needs to be scalable.}
    \label{field}
\end{figure}

With the received RF signal model established, we now consider the formulation of the RATF problem. We refer to a deployment scenario as a field, as shown in Figure \ref{field}. In each field, we want to develop a deep learning based authenticator for a specific set of transmitters $\{\mathcal{T}_i\}$. To develop such an authenticator, we need to collect training RF signals $Y$ captured on a set of receivers $\{\mathcal{R}_j\}$. When the authenticator is developed, it will be deployed on a set of receivers $\{\mathcal{R}_k\}$ different from $\{\mathcal{R}_j\}$.

From a practical perspective, the authenticator in a given field should be developed using a reasonable amount of training data and a viable training procedure. Furthermore, a developed authenticator should be scalably transferred to different receivers that were not used during the training procedure. Finally, the transferred authenticator should work correctly and identify the transmitters with reasonable accuracy on the newly deployed receivers. To summarize, we formulate the RATF problem, which takes the following aspects into consideration:
\begin{enumerate}
    \item Performance:
    When a developed authenticator is transferred to a new receiver, its performance on the new receiver is always the first metric to consider. In the task of transmitter fingerprinting, the performance is evaluated by the classification accuracy in a closed-set scenario. In an open-set scenario, besides the classification accuracy, the outlier detection rate is also considered to evaluate the performance.
    \item Practicality:
    The process of developing an authenticator needs to be practical in each given field. Collecting signals from multiple receivers is expensive and impractical in the field. Therefore, minimizing the number of training receivers $\{\mathcal{R}_j\}$ is essential in building a pratical solution.
    \item Cost of scalability:
    As shown in Figure \ref{field}, an authenticator trained on signals collected on receiver set $\{\mathcal{R}_j\}$ will be transferred and deployed on a different set of receivers $\{\mathcal{R}_k\}$. To reduce the cost of scalability, a transferred authenticator should not require any additional retraining or tuning, while still being able to work correctly.
\end{enumerate}

In this work, we consider the RATF problem in both closed-set and open-set scenarios. In both scenarios, we evaluate the performance of the approaches according the the above metrics. At the same time, we use only one receiver ($|\{\mathcal{R}_j\}|=1$) for field training, to enhance the practicality. Besides, an authenticator is directly applied on new receivers without further tuning to reduce the cost of scalability.

\section{Approach} \label{ap}

In this section, we first discuss the pre-processing of the received RF signals. Then, we propose our two-stage framework to solve the RATF problem. We present two deep learning based approaches that can be applied in the first stage of our framework and can improve the receiver-agnostic property of the calibrated feature extractor (FE). After that, we discuss the classifiers used in different scenarios. Finally, we provide details on the neural network architectures used in our approach.

\subsection{Pre-Processing} \label{preprocessing}
As modeled in Equation \eqref{signal_model}, an RF signal is impaired by three factors: the transmitter, the channel and the receiver. While this work focuses on removing receiver fingerprints in the transmitter fingerprinting problem, the effect of channel fingerprints cannot be ignored. Therefore, we need to account for the channel effects in the pre-processing step.

Given a received RF signal $y$, we first estimate the channel as $f_{H'}$, and then equalize the signals accordingly, as shown in Equation \eqref{estimate_model}:
\begin{equation}
\begin{split}
y' &= f_{H'^{-1}}(f_R(f_H(f_T(x)))) \\
   &= f_{R'}(f_T(x))
\end{split}
\label{estimate_model}
\end{equation}

It is shown in \cite{p4} that channel equalization can indeed significantly improve the transmitter fingerprinting performance. However, it is not a solution to receiver fingerprints, and different receivers for training and testing can still lead to a huge performance gap even in the presence of channel equalization, which is also shown in \cite{p4}. The reason for that may come from the different impairment models due to channel effects and receiver imperfections. The details of channel equalization steps are provided in Section \ref{ds}. We can safely equalize the signals since the technique is simple and only requires minimal amount of signal processing. It should be noted that we can perform channel equalization only on RF signals under certain protocols such as WiFi. For other protocols without training fields, such as Bluetooth, we can directly use raw IQ data without equalization, or refer to channel-agnostic fingerprinting techniques such as those mentioned in \cite{p5} and \cite{p12}.

\subsection{Two-Stage Framework: RXA} \label{3b}

After removing channel fingerprints in the pre-processing, we can consider the problem of receiver fingerprints. As shown in Equation \eqref{estimate_model}, we need to eliminate the receiver fingerprint $f_{R'}$ and estimate transmitter fingerprint $f_{T}$ from an equalized signal $y'$. Therefore, we propose an approach that can extract the transmitter-specific features $f_T$ from $y'$, regardless of the effects due to $f_{R'}$. However, due to the practical limitation of the RATF problem as mentioned in Section \ref{pf}, it is infeasible to collect a large-scale dataset in each targeted field. Therefore, we want to do an apriori calibration for transmitter-specific feature extractors.

\begin{figure*}[h]
    \centering
    \includegraphics[width=0.9\textwidth]{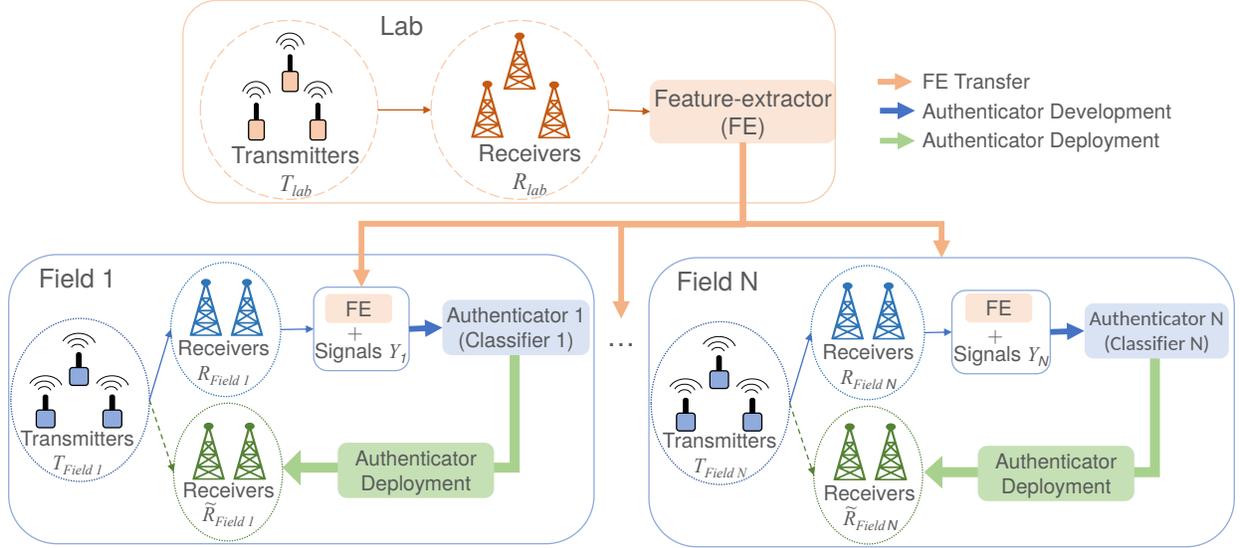}
    \caption{Demonstration of the two-stage learning process. The orange square shows Stage 1, feature-extractor (FE) calibration, which is performed on RF signals transmitted between transceivers $\mathcal{T}_{lab}$ and $\mathcal{R}_{lab}$. A calibrated feature-extractor (FE) is sent to each field, and each field can develop its own authenticator based on the transmitters to classify. A blue square shows Stage 2, classifier tuning, where each square represents a specific field. In each field, a classifier is developed based on both FE and RF signals collected on $\mathcal{R}_{field}$ and transmitted by $\mathcal{T}_{field}$. Finally, a trained classifier will be deployed on a different set of receivers $\mathcal{\Tilde{R}}_{field}$ to evaluate its performance.
    }
    \label{lab}
\end{figure*}

We propose a two-stage supervised learning framework as shown in Figure \ref{lab}. Our proposed framework aims to train receiver-agnostic (RXA) transmitter classifiers and is composed of the following two stages:
\begin{enumerate}
\item \textbf{Receiver-Agnostic Feature-Extractor Calibration}:
As stated in the problem formulation in the previous section, a field has a limitation in the number of receivers available for collecting the training signals for developing a transmitter authenticator due to practical reasons. Therefore, inspired by transfer learning, we can instead pre-train a transmitter feature-extractor (FE) before developing a field-specific transmitter authenticator for any field. For pre-training, we can have less limitation on the number of transceivers, so that the FE can leverage multiple receivers for learning. In other words, in the first stage, we calibrate a transmitter FE that is agnostic to receiver fingerprints in a controlled environment, represented by the lab in Figure \ref{lab}. The FE $f_{tx}$ needs to be trained on a large dataset, which includes signals from a large number of transmitters $\mathcal{T}_{lab}$ and received on multiple receivers $\mathcal{R}_{lab}$. $f_{tx}$ is trained to extract transmitter-specific features that are not impacted by receiver fingerprints. Once the FE is calibrated, we can use it in any targeted field as explained next.
\item \textbf{Classifier Training}: With the help of the pre-trained FE, each field can train its own authenticator with limited training signals. The authenticator in a field is trained on signals collected on receivers $\mathcal{R}_{field}$ and transmitted by transmitters $\mathcal{T}_{field}$. $\mathcal{R}_{field}$ has a practical limitation in its size and $\mathcal{T}_{field}$ is the specific set of transmitters to authenticate. Therefore, $|\mathcal{R}_{field}|$ is typically much smaller than $|\mathcal{R}_{lab}|$, and $|\mathcal{T}_{field}|$ is limited by the actual size of authentication task.
Within the scope of this work, an authenticator performs a classification task with or without the functionality of rejecting outliers. Therefore, we refer to the authenticator specifically as a classifier in the rest of the paper. Since the calibrated FE is agnostic to receiver fingerprints, a trained classifier, once developed, can be deployed on any unseen receiver without additional training. While each field needs to train its own classifier, the receiver-agnostic FE calibrated in the previous stage is not field-specific and can be reused as many times as necessary.
\end{enumerate}

As shown in Figure \ref{lab}, there are two different sets of transmitters and three different sets of receivers involved in the entire process, including both the training and deploying stages. In reality, the FE calibration stage is only performed once in the lab, and the calibrated FE can be sent to as many fields as needed. Therefore, it is natural to assume the transceivers deployed in fields are entirely different from the transceivers used in the lab. Hence, we consider the transceiver sets in different stages are mutually exclusive. In other words, we need to have $\mathcal{T}_{lab} \cap \mathcal{T}_{field} = \emptyset$, $\mathcal{R}_{lab} \cap \mathcal{R}_{field} = \emptyset$,  $\mathcal{R}_{field} \cap \mathcal{\Tilde{R}}_{field} = \emptyset$ and $\mathcal{R}_{lab} \cap \mathcal{\Tilde{R}}_{field} = \emptyset$.

In our proposed framework, we can see that the performance of a classifier depends on the quality of the FE, which is responsible for the receiver-agnostic property of the classifier. In particular, the design target of the FE is to obtain enough distinguishing capability among transmitters while being agnostic to the variances of receivers. Therefore, calibrating an effective FE in the first stage is essential, and the performance should be improved if we are able to enhance the receiver-agnostic property of the transmitter FE. In the following, we propose two approaches, which can be applied in the FE calibration stage to improve the receiver-agnostic property.

\subsection{SD-RXA} \label{3c}
\begin{figure}[h]
    \centering
    \includegraphics[width=0.45\textwidth]{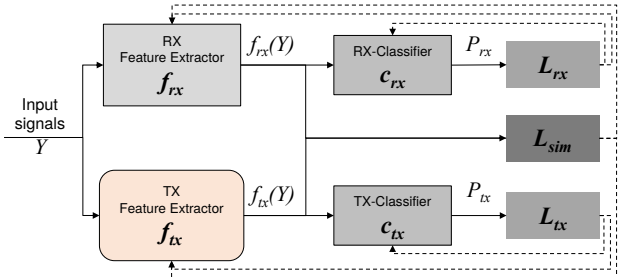}
    \caption{The SD-RXA approach architecture. A solid arrow represents the direction of a forward pass, and a dashed arrow represents the correspondence between a loss function and its targeted network for update through backpropagation. After the training converges, all the gray blocks are discarded. Only the orange-colored transmitter FE will be kept after the calibration.}
    \label{sd-RXA}
\end{figure}

Due to the asymmetry in receiver and transmitter features, it is natural to consider the receiver and transmitter feature distribution differences. Based on this intuition, we propose a statistical distance-based receiver-agnostic (SD-RXA) approach to improve the receiver-agnostic property in RXA. First, we assume the feature distributions of transmitters and receivers are uncorrelated. This assumption is based on the fact that receivers and transmitters have asymmetric features, and any signal capture should come from a random pair of transceivers. Based on this assumption, then, we extract the receiver features and transmitter features simultaneously for each captured RF signal. Finally, we optimize the FE so that the extracted receiver features and transmitter features tend to come from distributions with large statistical distance. The proposed architecture of SD-RXA is shown in Figure \ref{sd-RXA}.

As shown in Figure \ref{sd-RXA}, the SD-RXA approach has two parallel branches to process a single batch of input simultaneously. Both branches have the same structure: an FE is concatenated with a classifier. The top branch in Figure \ref{sd-RXA} extracts receiver features and classifies between receivers $\mathcal{R}_{lab}$, and the bottom branch in Figure \ref{sd-RXA} extracts transmitter features and classifies between transmitters $\mathcal{T}_{lab}$. Each branch is optimizing its corresponding FE and classifier by a crossentropy loss as presented in Equations \eqref{ce_tx} and \eqref{ce_rx},
\begin{equation}
L_{tx} = -\frac{1}{N} \sum_{i=1}^N \sum_{c=1}^{C_{tx}} y_{i, c}\log{\hat{y}_{i, c}}, 
\label{ce_tx}
\end{equation}
\begin{equation}
L_{rx} = -\frac{1}{N} \sum_{i=1}^N \sum_{c=1}^{C_{rx}} y_{i, c}\log{\hat{y}_{i, c}}, 
\label{ce_rx}
\end{equation}
where $L_{rx}$ and $L_{tx}$ correspond to the losses of receiver-branch and transmitter-branch, respectively. In the meanwhile, $N$ is the number of samples per mini-batch, $\hat{y}_{i, c}$ is the predicted probability for the $c$-th class of the $i$-th training signal in a mini-batch, and $y_{i, c}$ is the corresponding ground-truth label. $C_{tx}=|\{\mathcal{T}_{lab}\}|$ and $C_{rx}=|\{\mathcal{R}_{lab}\}|$ are the numbers of classes of transmitters and receivers, respectively. The crossentropy losses are applied to ensure the FEs are extracting effective receiver and transmitter features.

Beyond the two classifiers, we want to then measure and maximize the distance between the distributions of receiver features and transmitter features. In a practical realization, we can obtain the corresponding distributions for mini-batches of signals during training. To maximize such distance, we add an additional distance loss $L_{dist}$, which is optimized only aiming to update the two FEs. $L_{dist}$ measures the statistical distance between the batch distributions of receiver features and transmitter features. We use the correlation coefficient as the metric $L_{dist}$ as shown in Equation \eqref{dist_loss},
\begin{equation}
L_{dist} = -\frac{1}{T^2}\sum_{t1=1}^T\sum_{t2=1}^T
||A_{tx, t1, t2} - A_{rx, t1, t2}||_2^2
\label{dist_loss}
\end{equation}
where $T$ is the length of a feature vector.

As shown in Equation \eqref{dist_loss}, $L_{dist}$ is the average element-wise $L_2$ distance between $A_{rx}^{T \times T}$ and $A_{tx}^{T \times T}$. $A_{rx}^{T \times T}$ and $A_{tx}^{T \times T}$ are the batch self-covariance matrices of the receiver features and transmitter features, respectively. For example, $A_{tx}$ is a $T \times T$ self-covariance matrix of transmitter features, and $A_{tx, t1, t2}$ corresponds to the entry at $t1-th$ row and $t2-th$ column in matrix $A_{tx}$. A single self-covariance matrix $A^{T \times T}$ for a receiver or transmitter is obtained as shown in Equation \eqref{A},

\begin{equation}
A^{T \times T} = 
\begin{pmatrix}
cov_{h_1, h_1} & cov_{h_1, h_2} &\hdots & cov_{h_1, h_T}\\
cov_{h_2, h_1} & cov_{h_2, h_2} &\hdots & cov_{h_2, h_T}\\
\vdots & \vdots &\vdots &\vdots \\
cov_{h_T, h_1} & cov_{h_T, h_2} &\hdots & cov_{h_T, h_T}\\
\end{pmatrix}
\label{A}
\end{equation}
where $h=f(y)$ is a feature vector produced by its corresponding feature extractor.

Finally, the overall loss function of SD-RXA is the combination of the crossentropy losses and the distance losses described above, as presented in Equation \eqref{sd-loss},

\begin{equation}
L_{SD-RXA} = L_{tx} + \alpha*L_{rx} + \beta*L_{dist}
\label{sd-loss}
\end{equation}
where $\alpha$ and $\beta$ are the hyperparameters for the weightings of the different losses. $L_{tx}$ and $L_{rx}$ are the crossentropy losses as shown in Equations \eqref{ce_tx} and \eqref{ce_rx}, and $L_{dist}$ is the distance loss as shown in Equation \eqref{dist_loss}. It should be noted that while the two FEs are trained using the overall loss, $L_{dist}$ does not affect the two classifiers.

After completing the training procedure, the transmitter FE $f_{tx}$ is fixed and will be sent to different fields for deployment.

\subsection{GAN-RXA}
While the SD-RXA approach can effectively take the receiver features into account, we also discovered the following challenges while experimenting with it. First, measuring and maximizing the statistical distance between the distributions of receiver features and transmitter features is nontrivial and even tricky in some cases \cite{p27}. It is nontrivial because it is hard to measure the exact similarity between the receiver and transmitter features. It is also hard to decide what is an appropriate distance between the two distributions. Therefore, there does not exist a global optimum value for the total loss. Finally, we cannot explicitly evaluate the effectiveness of the receiver-agnostic property of the transmitter FE during the training process. As a result, we want to improve beyond SD-RXA approach and thus propose the GAN-RXA approach, which is inspired by Generative Adversarial Nets (GAN) \cite{p42}.

\begin{figure}[h]
    \centering
    \includegraphics[width=0.45\textwidth]{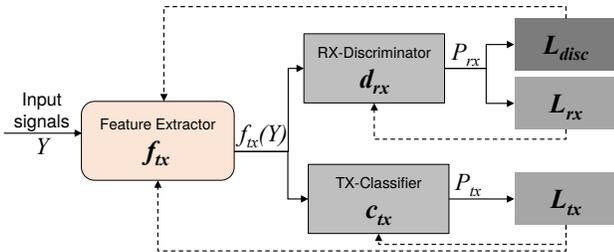}
    \caption{The architecture of the GAN-RXA approach. A solid arrow represents the direction of a forward pass, and a dashed arrow represents the correspondence between a loss function and its targeted component for update through backpropagation. Only the orange-colored FE will be kept after the calibration.}
    \label{gan-RXA}
\end{figure}

GAN-based models have been successfully applied in the area of RF signal recognition. For example, the authors in \cite{p60} have leveraged GAN to generate spoofing attacks. On the other hand, the authors in \cite{p58} built the Radio Frequency Adversarial Learning (RFAL) framework based on GAN, to identify rogue RF transmitters. In addition to attack-related scenarios, GAN has also been applied in device authentication to address practical limitations. For example, the authors in \cite{p59} developed an unsupervised GAN-based framework, called RFFE-InfoGAN (radio frequency fingerprint embedding information maximized GAN), to help device identification in noncooperative scenarios. In this work, the receiver variation is the practical limitation to be addressed, and the proposed GAN-RXA structure is shown in Figure \ref{gan-RXA}.

As shown in Figure \ref{gan-RXA}, the proposed GAN-RXA approach consists of 3 major components: the transmitter FE $f_{tx}$, a transmitter classifier $c_{tx}$ and a ``discriminator" $d_{rx}$. During the training stage, $c_{tx}$ learns to classify between the transmitters in the set $\mathcal{T}_{lab}$, and $d_{rx}$ learns to classify between the receivers in the set $\mathcal{R}_{lab}$. GAN-RXA takes each received RF signal $y$ as input and passes it through the FE $f_{tx}$ to extract feature vectors $f_{tx}(y)$. Then the feature vectors are fed into classifier $c_{tx}$ and the discriminator $d_{rx}$. $c_{tx}$ and $d_{rx}$ output vectors of probabilities $P_{tx}$ and $P_{rx}$, respectively. Each entry in $P_{tx}$ or $P_{rx}$ corresponds to the predicted probability for that transmitter or receiver. The intuition of this model is such that $f_{tx}$ learns to extract transmitter features that can be correctly classified by $c_{tx}$, and at the same time, the features cannot be correctly distinguished by $d_{rx}$. In this manner, the FE learns to extract transmitter-specific features without receiver-specific patterns. The loss functions $L_{tx}$ and $L_{rx}$ are crossentropy losses as shown in Equations \eqref{ce_tx} and \eqref{ce_rx}, respectively. The crossentropy losses are minimized to update $c_{tx}$ and $d_{rx}$, as well as $f_{tx}$. An additional loss $L_{disc}$ aims to remove the receiver-specific information embedded in the extracted transmitter-features $f_{tx}(Y)$, so that a well-trained discriminator fails to distinguish between receivers. $L_{disc}$ is shown in Equation \eqref{disc_loss},

\begin{equation}
L_{disc} = \frac{1}{NC_{rx}} \sum_{i=1}^N \sum_{c=1}^{C_{rx}} \{\exp{[(\hat{y}_{i, c} - z_{i, c})^2]} - 1 \}
\label{disc_loss}
\end{equation}
where $N$ is the number of samples per mini-batch, $C_{rx} = |\mathcal{R}_{lab}|$ is the number of receivers in the lab, $\hat{y}_{i, c}$ is the predicted probability for the $c$-th receiver $r_c \in \mathcal{R}_{lab}$ for the $i$-th training signal in a mini-batch, and $z_{i, c}$ is the ground-truth distribution of occurrence of $r_c$. The ground-truth distribution frequency of a specific receiver $r_c$ is its frequency of occurrence in the entire training set. 

Finally, the loss function $L_{GAN-RXA}$ is minimized only to update the FE as shown in Figure \ref{gan-RXA}, $L_{GAN-RXA}$ is shown in Equation \eqref{gan-loss},

\begin{equation}
L_{GAN-RXA} = L_{disc} + \gamma * L_{tx}
\label{gan-loss}
\end{equation}
where $\gamma$ is a hyperparameter coefficient to ensure the transmitter-distinguishability. Ideally, $f_{tx}$ should generate features so that $d_{rx}$ can only output random guesses for the receiver labels. However, in practice, if $L_{disc}$ updates $f_{tx}$ too aggressively, $f_{tx}$ loses the capability to distinguish between transmitters at the same time. Therefore, $L_{GAN-RXA}$ includes $L_{tx}$ additional to $L_{disc}$ with a weighting parameter $\gamma$, to serve as a constraint on the transmitter-distinguishability.

We conduct experiments with cross-validation to choose the value of $\gamma$. Through the experimentation, we find that a small $\gamma$ will cause the FE $f_{tx}$ to fail in learning the TX features, and a large $\gamma$ will cause $f_{tx}$ to learn RX-dependent features. We experimentally find $\gamma=1$ to balance these impacts, and use it throughout the experiments in Section \ref{results} for consistency.

The complete training of this approach is divided into two parts, where in the first part all three components $f_{tx}$, $d_{rx}$, and $c_{tx}$ are updated. Then, in the second part, only $f_{tx}$ and $d_{rx}$ are updated. The pseudo-code to train the model is provided below in Algorithm \ref{gan-RXA-alg}.
\begin{algorithm}
    \caption{GAN-RXA Update}
    \label{gan-RXA-alg}
    \renewcommand{\algorithmicrequire}{\textbf{Input:}}
    \renewcommand{\algorithmicensure}{\textbf{Output:}}
    \begin{algorithmic}[1]
        \REQUIRE mini-batch of received signals $x$
        \ENSURE transmitter-feature-extractor $f_{tx}$
        \STATE Initialize $f_{tx}$, $c_{tx}$ and $d_{rx}$ with random weights
        \WHILE{$L_{tx}$ is still decreasing}
            \FOR{$E_1$ epochs}
                \STATE freeze $d_{rx}$
                \STATE Train $f_{tx}$ and $c_{tx}$ with $L_{tx}$
            \ENDFOR
            \FOR{$E_2$ epochs}
                \STATE Freeze $f_{tx}$ and $c_{tx}$
                \STATE Train $d_{rx}$ with $L_{rx}$
            \ENDFOR
            \FOR{$E_3$ epochs}
                \STATE Freeze $d_{rx}$ and $c_{tx}$
                \STATE Train $f_{tx}$ with $L_{GAN-RXA}$
            \ENDFOR
        \ENDWHILE
        
        \STATE Freeze $c_{tx}$
        
        \WHILE{$d_{rx}$ is able to distinguish $\{\mathcal{R}_{lab}\}$}
            \IF{this training stage is looped over $L$ times}
                \STATE BREAK
            \ENDIF
            \FOR{$E_4$ epochs}
                \STATE Freeze $f_{tx}$
                \STATE Train $d_{rx}$ with $L_{rx}$
            \ENDFOR
            \FOR{$E_5$ epochs}
                \STATE Freeze $d_{rx}$
                \STATE Train $f_{tx}$ with $L_{GAN-RXA}$
            \ENDFOR
        \ENDWHILE

        \RETURN $f_{tx}$
    \end{algorithmic}
\end{algorithm}

It can be observed that in the first part, the whole structure is updated step by step, to achieve an optimal possible transmitter-distinguishability. Then, in the second part, the model focuses on removing the receiver fingerprints in $f_{tx}$, while retaining the power of distinguishing transmitters. The intuition behind the second part is such that $L_{GAN-RXA}$ is not yet optimized after the first part converges. It should be noted that the learning rate in the second part needs to be small enough so that the $f_{tx}$ is still able to extract effective transmitter features and $c_{tx}$ can still work properly.

\subsection{Classifiers} \label{cls}

After an FE is calibrated in the first learning stage, it is sent to every field which requires a classifier. In a given field, the possible classification tasks include both closed-set and open-set scenarios, the classifiers also have two architectures accordingly.

\begin{enumerate}
    \item Closed-set Classifier:
    In a closed-set scenario, a classifier needs to output predictions of transmitter classes. In fact, it performs the same task as a classifier and discriminator used in the FE calibration stage. Therefore, it should have the same architecture as them. To be more specific, a closed-set classifier has $C_{tx}$ output classes, where $C_{tx} = |\mathcal{T}_{field}|$ is the number of transmitters to be authenticated in the field. For each given input $y$, the classifier outputs a set of probabilities $P_{tx}$ for all classes, where $Pr(\hat{y}=c_i|y)$ is the predicted probability that $y$ is transmitted from transmitter $c_i$. Finally, the classifier decides the transmitter class $c_i$ which has the highest predicted probability, as shown in Equation \eqref{closedset}:
    \begin{equation}
    \max_{c_i} Pr(\hat{y}=c_i|y), \; \forall c_i \in \mathcal{T}_{field}
    \label{closedset}
\end{equation}

    \item Open-set Classifier:
    For open-set problems, the FE does not change, and we only need to slightly change the classifier in the field training stage. According to the results in \cite{p1}, one-versus-all (OVA) is the best performing classifier in open-set tasks. Therefore, we train a distinct OVA classifier for each class $c_i$, and each classifier has exactly two outputs for each input $y$: $\hat{y} = c_i$ and $\hat{y} \neq c_i$. Therefore, the classifier will have a set of probabilities $P_{tx}$. Similar to the closed-set scenario, $Pr(\hat{y}=c_i|y)$ is the predicted probability that $y$ is transmitted from transmitter $c_i$. Therefore, we will have $C_{tx}=|\mathcal{T}_{field}|$ classifiers, where $|\mathcal{T}_{field}|$ is the number of known transmitters in the field. Then, we reject a signal $y$ as outlier if the classifier fails to confidently choose a predicted class for it, with an experimentally chosen threshold $\tau$. Specifically, if the classifier yields probabilities as in Equation \eqref{openset}, it will decide the input $y$ as a signal from an outlier transmitter. $\tau$ is chosen so that an acceptable percentage of signals from known transmitters are classified as outliers. If a signal is not rejected, then it will be classified to belong to transmitter class $c_i$, which has the highest predicted probability, as shown in Equation \eqref{closedset}.
\end{enumerate}

\begin{equation}
    Pr(\hat{y}=c_i|y)<\tau, \; \forall c_i \in \mathcal{T}_{field}
    \label{openset}
\end{equation}

\subsection{Network Architecture}
We use modified ResNet-18 \cite{p43} as the backbone of the FE. The architecture is modified only to fit the input shape. Besides, a classifier or discriminator should be able to capture potential non-linearities within a feature vector while preventing overfitting. Therefore, without loss of generality, we use the same architecture, consisting of two concatenated dense layers, for both the classifier $c_{tx}$ and the discriminator $d_{rx}$.

\begin{figure}
    \centering
    \subfigure[Architecture of a feature-extractor (FE). The output is a feature vector of length $T$.]
    {\includegraphics[width=2.5cm]{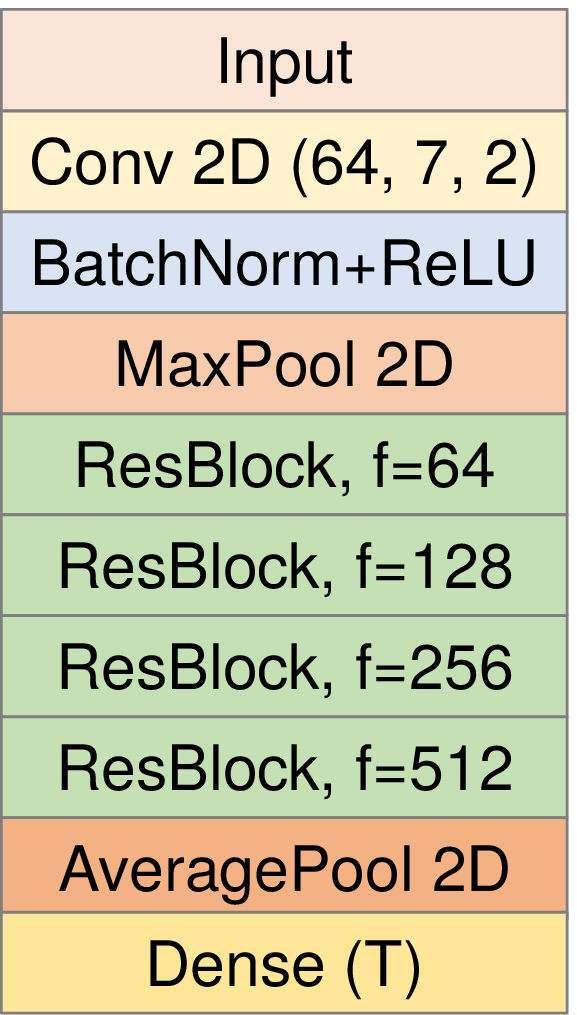}}
    \hfill
    \subfigure[\centering Architecture of a ResBlock with $f$ filters A ResBlock is a building block in an FE.]
    {\includegraphics[width=2.5cm]{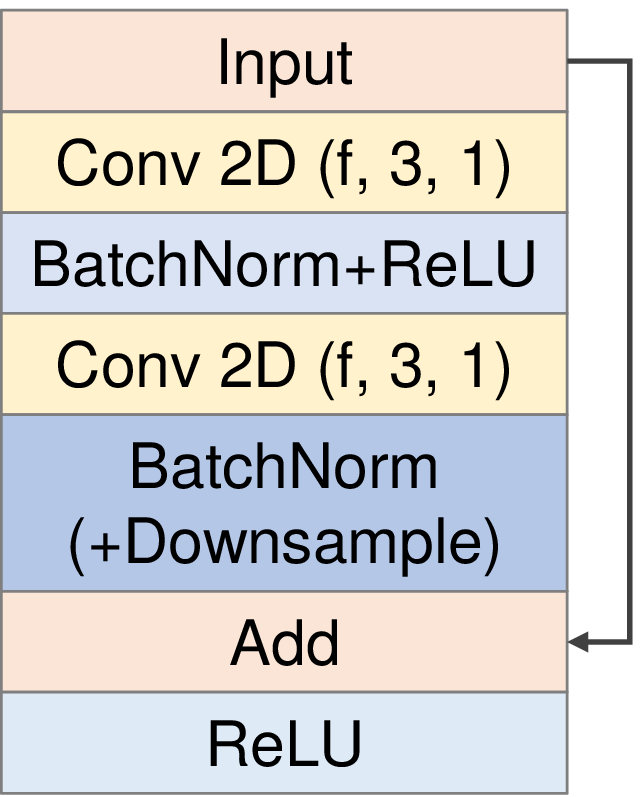}}
    \hfill
    \subfigure[\centering Architecture of a classifier and discriminator. The output size $n$ is number of classes to classify between depending on the specific task.]
    {\includegraphics[width=2.5cm]{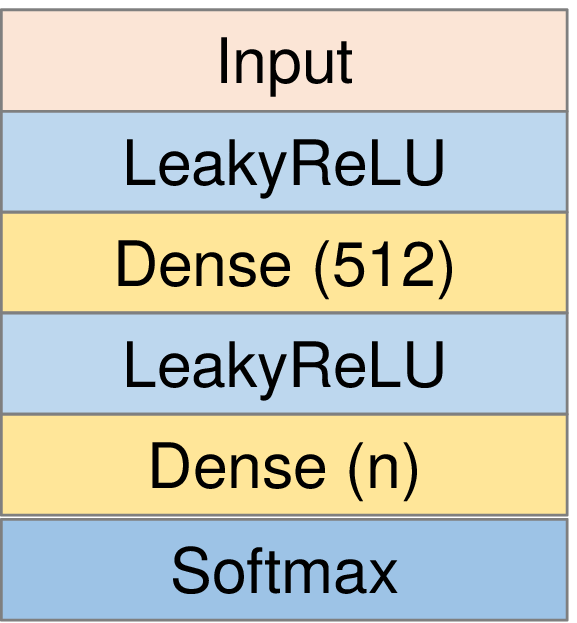}}
    \caption{The neural net architecture of each component in the proposed model. A notation $(f, s_1, s_2)$ represents $f$ filters of size $(s_1, s_2)$.}
    \label{nn}
\end{figure}

A detailed implementation of the model is illustrated in Figure \ref{nn}. To be more specific, the architecture of an FE is shown in Figure \ref{nn} (a), the architecture of a ResBlock, which is a building block in an FE is shown in   Figure \ref{nn} (b), and a classifier (discriminator) architecture is shown in  Figure \ref{nn} (c). For the classifier, the output classes vary according to its task, as explained in Section \ref{cls}. 

\section{Dataset}\label{ds}

WiSig is a large-scale WiFi dataset containing 10 million packets captured from 174 off-the-shelf WiFi transmitters and 41 USRP receivers over 4 captures spanning a month \cite{p4}. The size of the entire dataset is too large for efficient loading and processing, so we only use a subset from it in our experiments.

For the purpose of the experiments in this paper, we use over 2 million signals from 50 Atheros WiFi transmitters and 28 USRP receivers captured in 4 days. The transmitters include 49 AR5212/AR5213's and 1 AR9580, and the receivers include 3 B210's, 9 X310's and 16 N210's. The transceiver pairs are selected so that each receiver has successful receptions from at least 50\% of the transmitters. To eliminate the biases possibly introduced by the messages carried by a signal, we use exactly same partial preambles \cite{p44} in a WiFi packet for all signals, so that each signal contains the entire Legacy Long Training Field (L-LTF) and Legacy Short Training Field (L-STF) sampled at 25MHZ and has $400 \times 2$ IQ samples.

As mentioned in Section \ref{preprocessing}, the only pre-processing we perform on the raw data is channel equalization. It is worth noting that in this work, we use equalized data for all experiments, as it has been shown to outperform unequalized data \cite{p4}. To perform such channel equalization, we follow the procedure as in \cite{p4}, which is detailed as follows. First, we downsample a signal capture to change its sampling rate from 25 Msps to 20 Msps, which is the nominal sampling rate for WiFi, and apply autocorrelation on the L-STF to accurately detect a packet start. Then, we estimate the frequency offset and correct it accordingly. After that, we estimate and equalize the channel using minimum mean-square error (MMSE) on L-LTF. After this step, since the frequency offset can contribute to the fingerprints for both receivers and transmitters, we reapply it to the signal. Finally, we upsample the signals back to 25Msps. The signal processing for detection and channel estimation is applied using MATLAB R2019b WLAN toolbox with the default parameters.

\begin{table}
\caption{Dataset Separation}
\centering
  \begin{tabular}{|c|c|c|c|}
  \hline
        & $\mathcal{R}_{lab}$, $\mathcal{T}_{lab}$
        & $\mathcal{R}_{field}$, $\mathcal{T}_{field}$
        & $\mathcal{\Tilde{R}}_{field}$, $\mathcal{T}_{field}$\\
    \hline
     Day & Feature-Extractor &  & \\
     1, 2 & $f_{tx}$ Training &  & \\
    \hline
     Day & & Field Classifier & \\
     3 && Training &\\
    \hline
     Day & &  & Field Classifier\\
     4 &&& Testing \\
    \hline
  \end{tabular}
  \label{data}
\end{table}

We divide the dataset into 3 parts, as shown in Table \ref{data}. The training data and testing data come from different sets of transceivers and different days to minimize shared information between the training and testing signals. We use the signals from day 1 and day 2 for feature-extractor $f_{tx}$ training in the lab, signals from day 3 for operational field classifier training, and signals from day 4 for testing. The lab training dataset has the largest size to enhance receiver-agnostic property. The field training dataset only has limited size, with $|\mathcal{R}_{field}|=1$. $20\%$ signals randomly sampled from each training dataset are reserved for validation purpose.

\section{Experiments and Evaluations} \label{ee}
In this section, we first introduce the baseline approaches. Next, we evaluate the approaches according to the metrics of the RATF problem that were defined in Section \ref{pf}. Then, we evaluate the factors that can impact the quality of the FE. Finally, we consider the open-set problem and evaluate our proposed approaches as well as the baseline approaches for this problem.

For each experiment, we specify the number of receivers and transmitters in different sets of our framework as described in Section \ref{3b}, and then randomly select the transceiver sets. Each setting is repeated 5 times with different randomization, and the average performance is reported. Unless additionally specified, all approaches are trained and tested on exactly the same sets of transceivers and signals.

\subsection{Baselines}
\begin{enumerate}
    \item Naive approach: This approach is a simple transmitter classifier trained and tested in the field. It does not have the access to lab calibration. To ensure comparability, the architecture of this approach has the same FE and classifier structure as shown in Figures \ref{sd-RXA} and \ref{gan-RXA}, but only with the transmitter-branch. Although it has the FE, the FE is only trained in the field without any prior calibration, as it does not have access to the large training dataset. This approach is used to demonstrate the effectiveness of the proposed two-stage RXA framework.
    \item Exhaustive approach: This approach is similar to the naive approach. While it also has no access to the lab calibration stage, it has access to a considerably large set of receivers in the field. In other words, this approach is impractical in reality. This approach is used to demonstrate the practicality of the proposed approaches. 
    \item Basic-RXA: This approach has exactly the same structure as the naive approach: an FE together with a classifier. However, it has access to the entire calibration process in the lab. Therefore, its FE is pre-trained in the lab and delivered to fields. In other words, this approach incorporates the two-stage learning framework. While this approach itself can show the rationality of the proposed two-stage learning framework, it also serves as the baseline to demonstrate the effectiveness of the proposed SD-RXA and GAN-RXA approaches in their improved receiver-agnostic properties.
\end{enumerate}

\subsection{Performance Comparison}

\begin{table}
\caption{Metrics and Different Approaches Comparison}
\centering
  \begin{tabular}{|c|c|c|c|}
  \hline
        & \multirow{2}{*}{Practicality}
        & Cost of 
        & \multirow{2}{*}{Performance} \\
        &&Scalability&\\
    \hline
     Naive & \textbf{High} & \textbf{Low} & Bad  \\
    \hline
     Exhaustive & Low & \textbf{Low} & \textbf{Good} \\
    \hline
     RXA & \textbf{High} & \textbf{Low} & \textbf{Good} \\
     \hline
  \end{tabular}
  \label{metric}
\end{table}

To evaluate different approaches, other than accuracy, we also use the metrics as shown in Table \ref{metric}. As discussed in Section II, these aspects of an approach must be considered for realistic solutions. All the approaches considered here have a low cost of scalability, since none of them requires retraining on testing receivers. We can see that the exhaustive approach is impractical because it always require a large number of receivers in the field to train the classifier. On the other hand, while the naive approach is practical, it cannot yield classifiers that are transferable across receivers, since it does not take the receiver's fingerprints into account. To evaluate the performances, a quantitative and more detailed experimental comparison of the approaches is presented in the following.

\begin{figure}[h]
    \centering
    \includegraphics[width=0.45\textwidth]{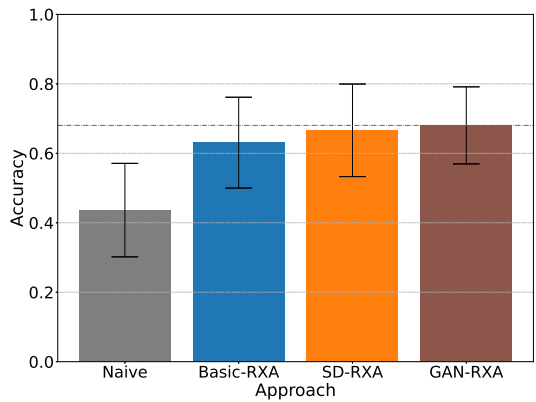}
    \caption{Comparison of proposed approaches vs baselines. All RXA approaches use $|\mathcal{R}_{lab}|=25$ and $|\mathcal{T}_{lab}|=40$. All approaches use $|\mathcal{R}_{field}|=1$, $|\mathcal{T}_{field}|=10$ and $|\mathcal{\tilde{R}}_{field}|=2$.} 
    \label{allap}
\end{figure}
In Figure \ref{allap}, we present the overall classification accuracy of the proposed approaches as well as the naive approach. We show that a calibrated FE brings $19.5\%$ improvement in field testing classification accuracy. This result demonstrates that the field transmitter classifier is essentially benefiting from the FE calibration stage in the RXA framework. As discussed in Section \ref{3b}, the diversity of transmitters and receivers in the FE calibration stage can help to improve the capability to recognize transmitter features of the classifiers, and thus increases the robustness against distortions due to receiver fingerprints. Beyond that, SD-RXA increases another $3.5\%$ classification accuracy. This result suggests the validity of the assumption on the asymmetry in receiver and transmitter features, which is proposed and discussed in Section \ref{3c}.

Meanwhile, GAN-RXA further boosts the accuracy by $1.5\%$, and achieves the best performance. Therefore, it is demonstrated that the proposed GAN-based network indeed helps the calibration of FE. Specifically, GAN-RXA improves the capability in suppressing the receiver-dependent features via our adversarial training procedure.

\begin{figure}[h]
    \centering
    \includegraphics[width=0.45\textwidth]{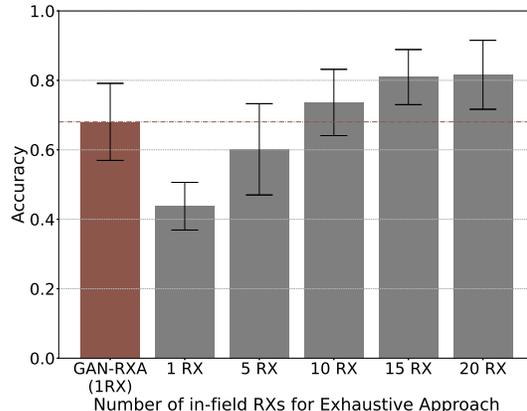}
    \caption{Comparison of proposed approach vs exhaustive approach with a different number of receivers (RXs). The GAN-RXA approach uses $|\mathcal{R}_{lab}|=25$, $|\mathcal{T}_{lab}|=40$ and $|\mathcal{R}_{field}|=1$. For the exhaustive approach,  different numbers of receivers are used in the field (see the X-axis of the figure). All approaches use $|\mathcal{T}_{field}|=10$ and $|\mathcal{\tilde{R}}_{field}|=2$.}
    \label{exhaustive}
\end{figure}
In Figure \ref{exhaustive}, we illustrate that GAN-RXA outperforms the exhaustive approach with 5 receivers and has a comparable performance with the exhaustive approach with 10 receivers. This result shows our proposed GAN-RXA approach significantly shrinks the number of receivers in the field to build an accurate transmitter classifier. While the exhaustive approach can have better classification accuracy when the number of in-field training receivers is large, it also becomes impractical in realistic scenarios. Therefore, the GAN-RXA approach makes the receiver-agnostic classifier a practical solution. Besides, the exhaustive approach reaches a saturation level in terms of classification accuracy with 15 receivers, and the performance stops improving.

In summary, the experiments show that the two-stage RXA framework brings substantial improvement to the performance of a transmitter classifier which is trying to be receiver-agnostic. Besides, it also demonstrates feasibility to solve the RATF problem due to its practicality and low cost of scalability. Finally, the GAN-RXA approach is shown to further improve the receiver-agnostic property of the calibrated FE and brings additional $5.0\%$ improvement compared to Basic-RXA in the field testing classification accuracy as shown in Figure \ref{allap}.

\subsection{Impacting factors}

As we have shown the superiority of the proposed approaches, in this subsection we consider how different experimental settings can affect the quality of the FE. At the same time, we want to assess the important factors in calibrating an effective FE. We vary $|\mathcal{T}_{lab}|$ and $|\mathcal{R}_{lab}|$ respectively and evaluate the results.We also investigate how $|\mathcal{T}_{field}|$ can affect classification accuracy. Finally, we look into the situation when the models of receivers in $\mathcal{\tilde{R}}_{field}$ are different from those in $\mathcal{R}_{lab}$ and $\mathcal{R}_{field}$.

\begin{figure}[h]
    \centering
    \includegraphics[width=0.45\textwidth]{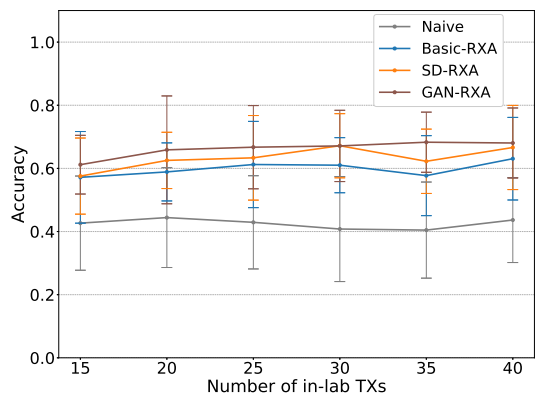}
    \caption{Comparison of $|\mathcal{T}_{lab}|$ vs performance.
    All RXA approaches use $|\mathcal{R}_{lab}|=25$ and varying $|\mathcal{T}_{lab}|$. All approaches use $|\mathcal{R}_{field}|=1$, $|\mathcal{T}_{field}|=10$ and $|\mathcal{\tilde{R}}_{field}|=2$.}
    \label{txs}
\end{figure}
\begin{figure}[h]
    \centering
    \includegraphics[width=0.45\textwidth]{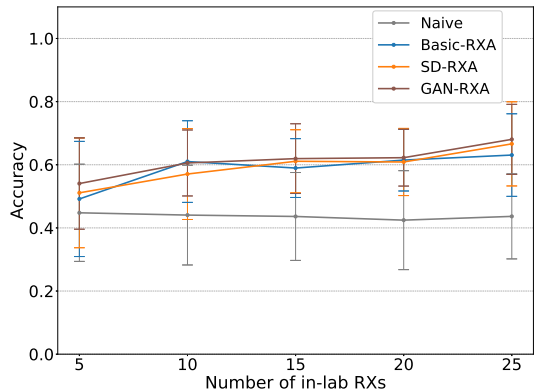}
    \caption{Comparison of $|\mathcal{R}_{lab}|$ vs performance. All RXA approaches use $|\mathcal{T}_{lab}|=40$ and varying $|\mathcal{R}_{lab}|$. All approaches use $|\mathcal{R}_{field}|=1$, $|\mathcal{T}_{field}|=10$ and $|\mathcal{\tilde{R}}_{field}|=2$.}
    \label{rxs}
\end{figure}
In Figure \ref{txs}, we conduct the experiments with a fixed number of in-lab receivers with a varying number of in-lab transmitters.
In contrast, in Figure \ref{rxs} we conduct the experiments with a fixed number of in-lab transmitters with a varying number of in-lab receivers. From both figures, we can observe that the performance of the naive approach is always fluctuating at a relatively low level and is always inferior to other approaches. Since the naive approach does not have the FE calibration stage, its performance indeed provides a baseline for each random combination of field transceivers. On the other hand, all three RXA approaches tend to improve as the number of transceivers increases in general. Besides, in both experiments, the GAN-RXA approach provides the highest testing classification accuracy for all the considered number of transceivers.

In the experiment which varies $|\mathcal{T}_{lab}|$, as shown in Figure \ref{txs}, GAN-RXA always has observable accuracy improvement compared to the Basic-RXA approach. The second best approach is the SD-RXA approach, and its performance is upper bounded by GAN-RXA and lower bounded by Basic-RXA. Its instability in performance might be the result of suboptimal statistical distance between the transmitter and receiver feature distributions in some cases.

In the experiment which varies $|\mathcal{R}_{lab}|$, as shown in Figure \ref{rxs}, GAN-RXA still has the highest classification accuracy in all experiments. When $|\mathcal{T}_{lab}|=10$ and $|\mathcal{T}_{lab}|=15$, we can see that GAN-RXA has similar accuracy as Basic-RXA. However, GAN-RXA still has smaller variances across different randomizations. Therefore, we can conclude our proposed GAN-RXA approach demonstrates the best stability across different experiments as well. 

\begin{figure}[h]
    \centering
    \includegraphics[width=0.45\textwidth]{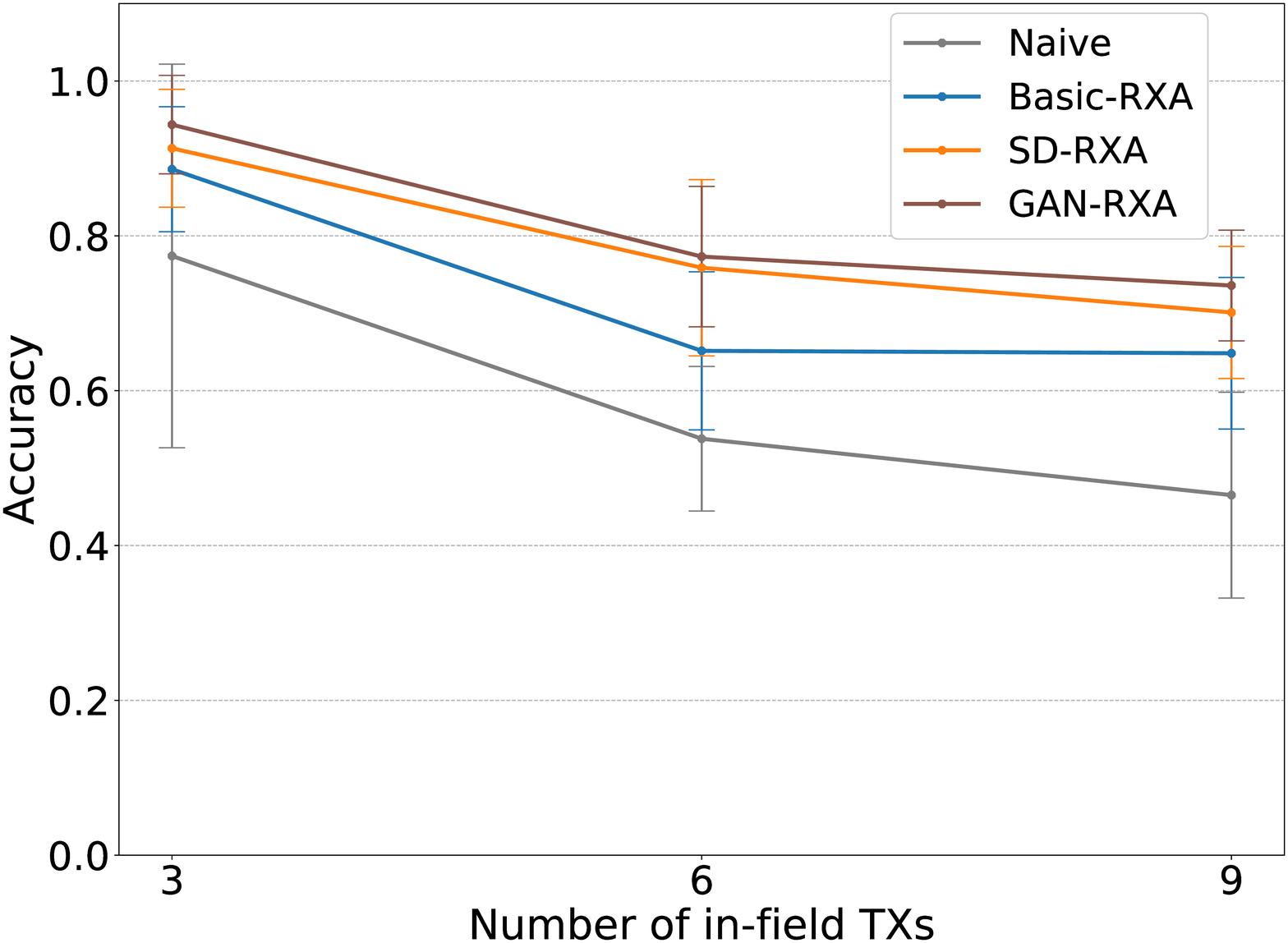}
    \caption{Comparison of $|\mathcal{T}_{field}|$ vs performance. All RXA approaches use $|\mathcal{R}_{lab}|=25$ and $|\mathcal{T}_{lab}|=40$. All approaches use $|\mathcal{R}_{field}|=1$, $|\mathcal{T}_{field}|=10$ and $|\mathcal{\tilde{R}}_{field}|=2$.}
    \label{txs-field}
\end{figure}

In Figure \ref{txs-field}, we conduct the experiments with a fixed number of in-lab transceivers with a varying number of in-field transmitters. We observe that the performances of all 3 approaches degrade as the number of transmitters in the field increases. This is a natural trend since a classifier is more likely to have a wrong decision on an input when there are more transmitters to be classified. The GAN-RXA outperforms all other approaches again in this experiment, achieving $94.4\%$ accuracy in small-scale in-field transmitter settings.

\begin{figure}[h]
    \centering    \includegraphics[width=0.45\textwidth]{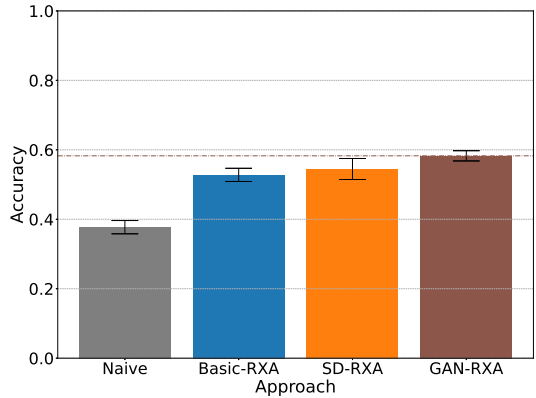}
    \caption{Evaluation of performance when the models of $\mathcal{\tilde{R}}_{field}$ are not included in $\mathcal{R}_{lab}$ and $\mathcal{R}_{field}$. All RXA approaches use $|\mathcal{R}_{lab}|=18$ and $|\mathcal{T}_{lab}|=40$. All approaches use $|\mathcal{R}_{field}|=1$, $|\mathcal{T}_{field}|=10$ and $|\mathcal{\tilde{R}}_{field}|=2$.}
    \label{diffrx}
\end{figure}

Finally, we consider the case where the model of testing receiver has never been seen in any of the training stages. In this experiment, $\mathcal{R}_{lab}$ consists of 15 N210 and 3 B210 receivers, $\mathcal{R}_{field}$ has one N210 receiver, and $\mathcal{\tilde{R}}_{field}$ has 2 X310 receivers. The experiment results are presented in Figure \ref{diffrx}. From the plot, we can observe that all the approaches have slightly lower classification accuracy than Figure \ref{allap} when the models of testing receivers were used in the training stage. However, our proposed RXA framework still helps to improve the performance over the naive approach. Moreover, the SD-RXA has another slight improvement over the Basic-RXA approach, and GAN-RXA still has the highest classification accuracy. This result shows that different receiver models in the training and testing stages will negatively impact classification accuracy. However, our proposed RXA framework can still work reasonably well in such scenarios. 
Finally, this result also suggests that even with a receiver-agnostic transmitter fingerprinting approach, prior knowledge about the deployment receiver type can be leveraged by adding such receiver type to the training set.

\subsection{Openset Tasks} \label{ops}
Openset problems can provide different insights from closed-set problems. It should be noticed that the open-set classifier performs both authentication and classification tasks at the same time, as explained in Section \ref{cls}. To be more specific, given an input signal $x$, the open-set classifier can either reject $x$ or accept and classify $x$. Therefore, we investigate the open-set problem in this section.

\begin{figure}[h]
    \centering
    \includegraphics[width=0.45\textwidth]{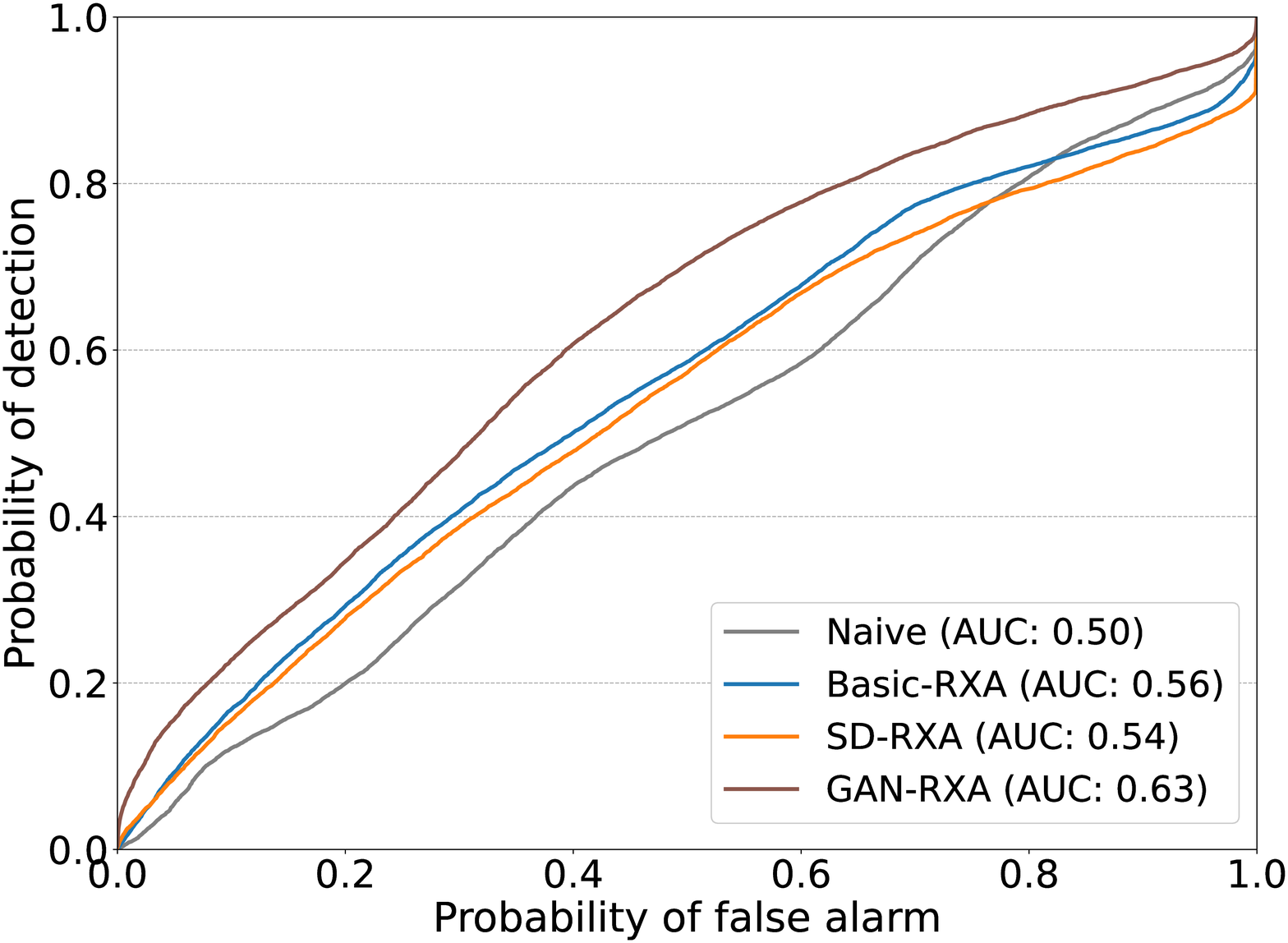}
    \caption{Openset outlier detection: ROC curves of anomaly detection for different approaches. All RXA approaches use $|\mathcal{R}_{lab}|=25$ and $|\mathcal{T}_{lab}|=40$. All approaches use$|\mathcal{R}_{field}|=1$ and $|\mathcal{\tilde{R}}_{field}|=2$. $|\mathcal{T}_{field}|=6$ is used as known set for training, and 4 different transmitters are introduced in the testing phase as outliers.}
    \label{op1}
\end{figure}

The first important task in open-set problems is outlier detection. In Figure \ref{op1}, we examine this aspect and present the ROC curves for different approaches. In the experiment, an outlier is a signal coming from a transmitter that is not seen during the training stage. Therefore, a true positive is a correct rejection of an outlier signal, and a false positive is a mistaken rejection of an authenticated signal. In Figure \ref{op1}, the probability of false alarm is the false positive rate, and the probability of detection is the true positive rate. Unfortunately, all approaches do not perform well in this task. This is primarily due to the fact that the distribution of transmitter features has vastly changed due to different operating receivers. The naive approach has the worst performance with a $0.50$ area under its ROC curve, which is equivalent to a random classifier. The Basic-RXA and SD-RXA have similar performances which are slightly better than that of the naive approach. The improvement can be attributed to the proposed RXA framework. Finally, GAN-RXA still performs best and yields a $0.63$ area under its ROC curve. Although this performance is not perfect, it does provide a substantial improvement over other approaches. This result demonstrates the effectiveness of the proposed GAN-RXA approach in the outlier detection task. At the same time, it also shows that the RATF problem is more challenging in the open-set scenario than in the closed-set scenario. In the open-set scenario, the receiver fingerprints cause distortions to the transmitter feature distributions, and thus introduce more ambiguity when there exist unseen transmitters with unknown feature distributions.

\begin{figure}[h]
    \centering
    \includegraphics[width=0.45\textwidth]{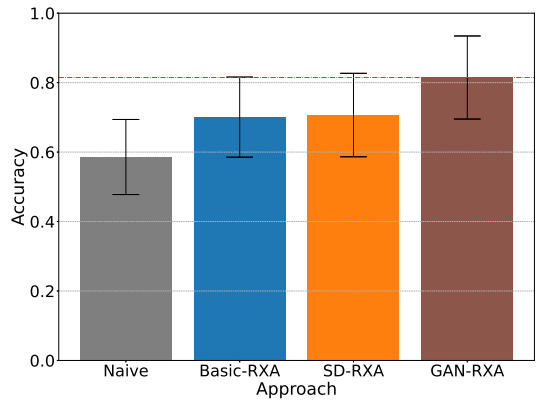}
    \caption{Openset transmitter classification: classification accuracy among correctly accepted transmitters for different approaches. All RXA approaches use $|\mathcal{R}_{lab}|=25$ and $|\mathcal{T}_{lab}|=40$. All approaches use $|\mathcal{R}_{field}|=1$ and $|\mathcal{\tilde{R}}_{field}|=2$. $|\mathcal{T}_{field}|=6$ is used as known set for training, and 4 different transmitters are introduced in the testing phase as outliers. We use the probability threshold $\tau$ as the decision boundary so that all approaches have approximately $15\%$ signals from the known transmitter set $\mathcal{T}_{field}$ are rejected.}
    \label{op2}
\end{figure}

Beyond the outlier detection task, the classification accuracy in authenticated sets is also important. Therefore, we continue the experiment and find average classification accuracy among the correctly accepted transmitters, and the result is shown in Figure \ref{op2}. In Figure \ref{op2}, we set the false alarm rate to be $15\%$ by experimentally finding the threshold $\tau$ in Equation \eqref{openset}, which means we will wrongly reject $15\%$ authenticated signals. We can observe the Basic-RXA and SD-RXA approaches have similar performance again. Besides that, the RXA-approaches still outperform the naive approach significantly, and GAN-RXA demonstrates much higher accuracy than all other approaches.

For all the approaches evaluated, we can observe performance improvement from Figure \ref{txs-field} (when $|\mathcal{T}_{field}|=6$) to Figure \ref{op2}. The reason for this improvement could be the rejection of the uncertain signals in the open-set scenario. Therefore, this result also suggests that we can improve the closed-set classification accuracy by rejecting uncertain signals.
 In this experiment, we can observe that while the Basic-RXA and SD-RXA approaches perform similarly, the GAN-RXA approach significantly outperforms other approaches. Therefore, we can conclude that the RXA framework can improve the performance in open-set task, while SD-RXA does not bring further improvement due to its instability in modeling the feature distributions. Furthermore, GAN-RXA remains the best choice among all considered approaches in the open-set problems.

The overall performance comparison among the approaches are presented in Table \ref{results}. We can observe that our proposed RXA approach can always improve the performance over the naive approach, while GAN-RXA outperforms the other approaches in all occasions.

\begin{table}
\caption{Results}
\centering
  \begin{tabular}{|c|c|c|c|}
  \hline
        & Closed-set
        & {Outlier Detection}
        & {Open-set Accuracy}\\

        & Accuracy
        & {ROC AUC}
        & {(15\% False Alarm)} \\
    \hline
     Naive & 43.6\% & {0.50} &
     {58.6\%}  \\
    \hline
     Basic-RXA & 63.1\% & {0.56}
     & {70.1\%}   \\
    \hline
     SD-RXA & 66.6\% & {0.54}
     & {70.7\%}  \\
     \hline
     GAN-RXA & \textbf{68.1\%} 
     & {\textbf{0.63}}
     & {\textbf{81.5\%} } \\
    \hline
  \end{tabular}
  \label{results}
\end{table}

\section{Conclusion} \label{conclusion}
In this work, we addressed the problem of receiver effects in the transmitter fingerprinting problems. We first formulated the RATF problem. We stress that in the problem of transferring classifiers across different receivers, while transmitter classification performance is the basic requirement, practicality and cost of scalability are two important metrics in evaluating a solution. Then, we proposed a practical and scalable two-stage learning framework (RXA) to address the RATF problem. We further proposed two deep-learning-based approaches, SD-RXA and GAN-RXA, to improve the receiver-agnostic property of the RXA framework. We evaluated the proposed approaches under different use cases, which involve both closed-set and open-set authentication scenarios. We showed that the proposed RXA approach can bring substantial improvement to both tasks, and our proposed GAN-RXA approach can further enhance the performance on top of Basic-RXA.

In the closed-set scenario, the two-stage learning framework improved the classification accuracy by $19.5\%$ compared to the naive approach, and GAN-RXA increases it by another $5.0\%$ compared to Basic-RXA. In summary, in a closed set problem, GAN-RXA brought a total improvement of $24.5\%$ over the naive approach. In the openset scenario, the two-stage learning framework improved the outlier detection ROC AUC by $12.0\%$, and also improved the classification accuracy on correctly accepted transmitters by $11.5\%$ compared to the naive approach. The GAN-RXA approach additionally improved the outlier detection ROC AUC by $12.5\%$ and the classification accuracy by $11.4\%$ compared to Basic-RXA. In summary, in openset problem, GAN-RXA brought a total improvement of $26.0\%$ in outlier detection ROC AUC and $22.9\%$ in classification accuracy over the naive approach. In both cases, we have demonstrated the effectiveness of the proposed approaches.

While our proposed approach has significantly improved the receiver-agnostic classification performance, it is still not as good as a classifier trained and tested on the same receiver \cite{p4}. One direction for future work is the enhancement of the receiver-agnostic feature extractors. In addition, it would be valuable to understand how the GAN-RXA deep learning model works. An interpretable deep learning model can help in improving the overall goals of transmitter fingerprinting. Another direction for future work is more investigation into the open-set problem as our current open-set performance is subpar. Finally, recognizing unauthorized transmitters and tracking their behavior can be potentially beneficial for security purposes. Hence, we will also look into this interesting but challenging problem and enhance the capabilities of the open-set framework.

\section*{Acknowledgments}
We thank Dr. Hazem Sallouha (KU Leuven, Belgium) for his insightful comments and suggestion throughout this paper.

\bibliographystyle{ieeetr}
\bibliography{ref}

\vfill

\end{document}